# An AI dose engine for fast carbon ion treatment planning.

**A. Quarz[1,2], A. De Gregorio[3], G. Franciosini[4,5], A. Schiavi[4,5], Z. Perkó[6], L. Volz[1], C. Hoog Antink[2], V. Patera[4,5] , M. Durante[1,7, 8] and C. Graeff[1,2]**

[1]GSI Helmholtzzentrum für Schwerionenforschung, Biophysics Department, Darmstadt, Germany
[2]Technische Universität Darmstadt, Department of Electrical Engineering and Information Technology, Darmstadt, Germany
[3]Dipartimento di Fisica, Sapienza Università di Roma, Italy
[4]Dipartimento di Scienze di Base e Applicate per l'Ingegneria, Sapienza Universita` di Roma, Roma, Italy
[5]INFN (Istituto Nazionale di Fisica Nucleare) section of Roma 1, Rome, Italy
[6]Delft University of Technology, Department of Radiation Science and Technology, Delft, The Netherlands
[7]Technische Universität Darmstadt, Department of Condensed Matter Physics, Darmstadt, Germany
[8]University Federico II, Department of Physics "Ettore Pancini", Naples, Italy
E-mail: c.graeff@gsi.de

## Abstract

*Purpose*

Monte Carlo (MC) simulations provide gold-standard accuracy for carbon ion therapy dose calculations but are computationally intensive, limiting their use in adaptive workflows. Analytical pencil beam algorithms offer speed but reduced accuracy in heterogeneous tissues. This study develops the first AI-based dose engine capable of predicting relative biological effectiveness (RBE)-weighted doses. Absorbed dose, α, and β parameters for optimisation are calculated at MC-level accuracy with a drastically reduced computational time.

*Materials and Methods*

We extended the transformer-based DoTA architecture to predict absorbed dose (C-DoTA-d), α (C-DoTA-α), and β (C-DoTA-β), introducing a cross-attention mechanism for α and β to combine dose and energy inputs. The training dataset consisted of approximately 70,000 pencil beams from 187 head-and-neck patients, with ground-truth values obtained using the GPU-accelerated MC toolkit FRED. Performance was evaluated on an independent test set using gamma pass rate (1%/1 mm), depth-dose, and isodose contour Dice coefficients. MC dropout–based uncertainty analysis was performed.

*Results*

Median gamma pass rates exceeded 98% for all predictions (99.76% for dose, 99.14% for α, 98.74% for β), with minima above 85% in the most heterogeneous anatomies. The Dice coefficient was 0.95 for 1% isodose contours, with slightly reduced agreement in high-gradient regions. Compared to MC FRED, inference was over 400× faster (0.032 s vs. 14 s per pencil beam) while maintaining accuracy. Uncertainty analysis showed high stability, with mean standard deviations below 0.5% for all models.

*Conclusions*

This AI-based dose engine achieves MC-quality predictions of absorbed dose and RBE model parameters in ~30 milliseconds per beamlet. Its speed and accuracy support online adaptive planning, paving the way for more effective carbon ion therapy workflows. Future work will expand to additional anatomical sites, beam geometries, and clinical beamlines.

# 1. Introduction

Particle therapy beams exhibit a steep maximum at the end of their range in the patient, known as the Bragg peak. Calculating this sharp dose gradient in heterogeneous tissue is particularly complex. For carbon ion therapy, the ions' variable relative biological effectiveness (RBE) - which also shows a maximum at the Bragg peak location - introduces a further challenge. Accurate dose calculation in carbon ion therapy, therefore, not only requires a precise dose model, but it also relies on the correct calculation of biological factors that depend on the spectrum of primary and secondary particles along the beam path in the patient [1,2].

Monte Carlo (MC) particle transport simulations are generally considered the gold standard for dose calculation and radiobiological modelling [3]. By considering the physics processes involved in the beam interaction with the patient, MC simulations accurately produce the lateral broadening and range mixing of the pencil beams (PBs). Recent advances in GPU-accelerated MC codes have allowed increasing introduction of MC dose engines in commercial treatment planning systems for proton therapy [4–6]. However, the increased complexity of particle transport for carbon ions compared to protons – requiring the simulation of multiple fragment species along the primary ion path and the associated computational expense still render MC carbon dose engines impractical for clinical routine. Computation time for even the fastest currently available carbon ion GPU MC codes is a significant bottleneck, particularly in scenarios requiring many or rapid plan optimisations, such as for robust 4D optimisation [7] and adaptive treatment workflows [8,9].

Due to this limitation, all carbon radiotherapy treatments are still planned using variations of analytical pencil beam algorithms (PBAs) [10,11]. However, these algorithms make inherent assumptions that drastically limit their accuracy in heterogeneous tissues [12]. Several studies have compared PBAs with MC methods, highlighting trade-offs between speed and precision in proton [13] and carbon [13,14] therapy. For high-precision treatment planning that fully exploits the accuracy available with carbon ion beams, MC-quality dose engines are essential.

For proton dose calculation, artificial intelligence (AI) has been successfully applied to achieve MC accuracy at massively reduced runtimes, sometimes even outperforming traditional PBA speed. Approaches such as DoTA [15], LSTM-based dose prediction [16,17], and DiscoGAN [18] have demonstrated the feasibility of rapid dose calculations of single PBs as well as for the full dose distribution [19]. Recent studies have also explored the use of AI to enhance the quality of PBA for protons [20,21].

The greater complexity of carbon ions due to nuclear fragmentation and their effect on the resulting RBE parameters present a challenge that has not been considered in any of the available AI dose engines yet. As any carbon therapy treatment planning system (TPS) needs to optimise the RBE-weighted dose, carbon ion dose engines must include the relevant biological parameters used in the RBE-weighted dose calculations, namely the alpha and beta parameters in the linear quadratic model of cell survival [11].

Previous works in carbon ion dose prediction [22,23] have focused solely on physical dose, omitting the biologically crucial RBE component. To our knowledge, this study represents the first attempt to fill in this gap by developing an AI-driven dose engine capable of jointly predicting physical dose, and $\alpha$ and $\beta$ parameters for RBE-weighted optimisation.

We extended the transformer-based DoTA [15] model, originally developed for proton physical dose calculation, and trained three models separately to predict the input parameters for the local effect model (LEM I) [24]: namely, the absorbed dose (C-DoTA-$d$), alpha (C-DoTA-$\alpha$) and beta (C-DoTA-$\beta$). Our model operates on a per-PB basis, enabling both optimisation and forward dose calculation.

In this proof-of-concept demonstration, we assess our model's feasibility on a cohort of head and neck cancer patients formerly treated within a carbon ion therapy pilot study at GSI (Darmstadt, Germany). As a baseline for comparison and to generate training data, we utilise the GPU MC toolkit FRED [4,25], which can score the alpha and beta parameters alongside the physical dose at fast per-pencil-beam calculation speeds of only a few seconds. The models deliver all three input parameters in only 32 ms, with a median gamma pass rate (GPR) of a gamma analysis with 1%/1 mm criteria and a 10% threshold >99%.

# 2. Materials and methods

## *2.1 Patient data*

Treatment planning data were taken from the GSI carbon ion pilot project for a cohort of head and neck cancer patients treated between 1997 and 2008 at GSI with carbon ions. The pilot project was approved by the ethical committee of the University of Heidelberg in 1997. Anonymised treatment plans of all patients are stored for research purposes at GSI. Informed consent is waived by the ethical committee of the University of Heidelberg for anonymised plans used for research purposes.

For this study, 225 head-and-neck cancer patients were randomly selected from the database. 187 patients' CTs were used for training and validation of the models, and 38 for the final test. For each patient, the target contours and isocentres of the clinical plans were reused for PB data generation, but to augment the data, additional beam angles were added.

## *2.2 Data generation*

As most patients were treated with lateral opposing fields from GSI's fixed horizontal beam line, also the immobilization masks were created for these angles. Therefore, for each patient, five positive and five negative couch angles were randomly selected from the range +/-[85°,105°]. Treatment fields with a PB spacing of 10 mm were generated in TRiP98 for these angles, and a BEV CT was generated for each of the unique PB coordinates.

All BEV CTs were set to an isotropic resolution of 1 mm, with lateral slices of 48x48 pixels centred on the PB coordinate. BEV CT slices extended through the entire CT. Proximal slices containing only air (HU<900) were removed, see Figure S1. BEV CTs that contained part of the shoulder or mask mounting structure were excluded.

Beam energy values in the dataset were restricted to the range 115 to 260 MeV/u to ensure a homogeneous representation of energies. For each BEV CT, two PB are selected, with an energy randomly sampled from the available discrete beam energies. If within this energy range, no Bragg Peak would be placed inside the target, this BEV CT was excluded. To achieve an even distribution, energies are excluded from sampling if the number of PB for this energy exceeds 99% of the maximum number of PB for any energy. If the Bragg Peak of the selected PB would be placed outside the patient, the energy was resampled. The final distribution of energies is shown in Figure S2.

The data generation resulted in 34706 and 1769 BEV-CTs for training and validation, respectively (corresponding to a 95%/5% split), with 2 PB of different energy each. For the test dataset, from the remaining 38 patients 6963 BEV CTs with 13926 PBs were generated. For each PB, ground truth (GT) data was computed with the GPU-based MC tool FRED [4,25].

## 2.3 FRED

A fast MC particle transport software was needed to compute the large amount of required GT data. We chose FRED [4,25], which is capable of GPU-accelerated dose calculations for protons and light ions. This work relies on the recently implemented carbon ion extension of FRED [4,25].

The simulation for carbon ions considers three main physics blocks: ionisation energy loss, multiple scattering (both implemented analogously to the proton transport models of FRED) and a phenomenological fragmentation model. The latter was developed directly from published fragmentation data and evaluated against this data as well as FLUKA simulations. Full details on the physics processes in the simulation are provided in the overview paper by De Simoni et al [4]. In addition to the absorbed dose, FRED supports RBE-weighted dose calculation with multiple scoring options. In this work, we relied on table-based RBE calculation using tabulated alpha and beta values as a function of energy for the first eight light ion species (up to Oxygen) generated with LEM I. Internally in FRED, the effect of mixed ion fields is resolved by dose averaging [11,26], in the same way as in TRiP98:

$$alpha = \frac{\sum_{i=1}^{n} a_i D_i}{\sum_{i=1}^{n} D_i} \tag{0.1}$$

$$beta = \left( \frac{\sum_{i=1}^{n} \sqrt{\beta_i} D_i}{\sum_{i=1}^{n} D_i} \right)^2 \tag{0.2}$$

Where $\alpha_i$ and $\beta_i$ are the energy-specific alpha and beta values for a specific ion type, n is the total number of ion species considered. $D_i$ is the dose contribution from each ion species. The mixed alpha and beta values were saved together with the absorbed dose. In the TRiP98 treatment planning system, these values are used alongside the physical dose to calculate the RBE-weighted dose. Hence, they present the target values for the C-DoTA-$\alpha$ and C-DoTA-$\beta$ models, respectively, while the absorbed dose output was the target for the C-DoTA-d model.

The simulation geometry consisted of the interpolated BEV CT, positioned so that its proximal entrance was located at the isocentre. The beam nozzle was simulated as a PMMA slab of 4mm thickness, positioned 10 cm in front of the BEV CT in beam direction. The beam was generated 20 cm in front of the isocentre. Each PB was simulated as a Gaussian beam with full width half maximum (FWHM) according to the CNAO beam characteristics at the respective PB energy. For all PBs, a momentum spread of 0.5% was assumed. Note that this does not reflect the characteristics of the CNAO beamline. A fully realistic simulation would introduce slight differences in the form of the beam's lateral spread (wider due to air drift) and range straggling (slightly broader). The simplified beam geometry was chosen to permit efficient training data generation. Each PB simulation used $10^7$ primaries for high-quality GT data. Dose, alpha, and beta were scored on the same grid as the BEV CT. FRED provides the dose in Gy/primary.

## 2.4 Model architecture

Figure 1 represents the C-DoTA architectures. For physical dose prediction, the vanilla DoTA architecture proved optimal for carbon ions, utilising normalised beam's-eye-view (BEV) CT scans and energy information as inputs (C-DoTA-d network). However, for α and β predictions, we extended the model by incorporating:

1. An additional convolutional encoder and transformer block processing both energy data and the physical dose output from the C-DoTA-*d* network (using a 1% dose threshold below which doses were zeroed out),

2. A cross-attention block to merge the BEV CT and dose information streams, and

3. A final decoder block for prediction output. The resulting models are called C-DoTA-α and C-DoTA-β, respectively.

Other model design choices in the original DoTA, such as the choice of convolutional and transformer block setups, hyperparameters (including learning rate schedules) and the optimiser configuration, were found to be optimal for this study.

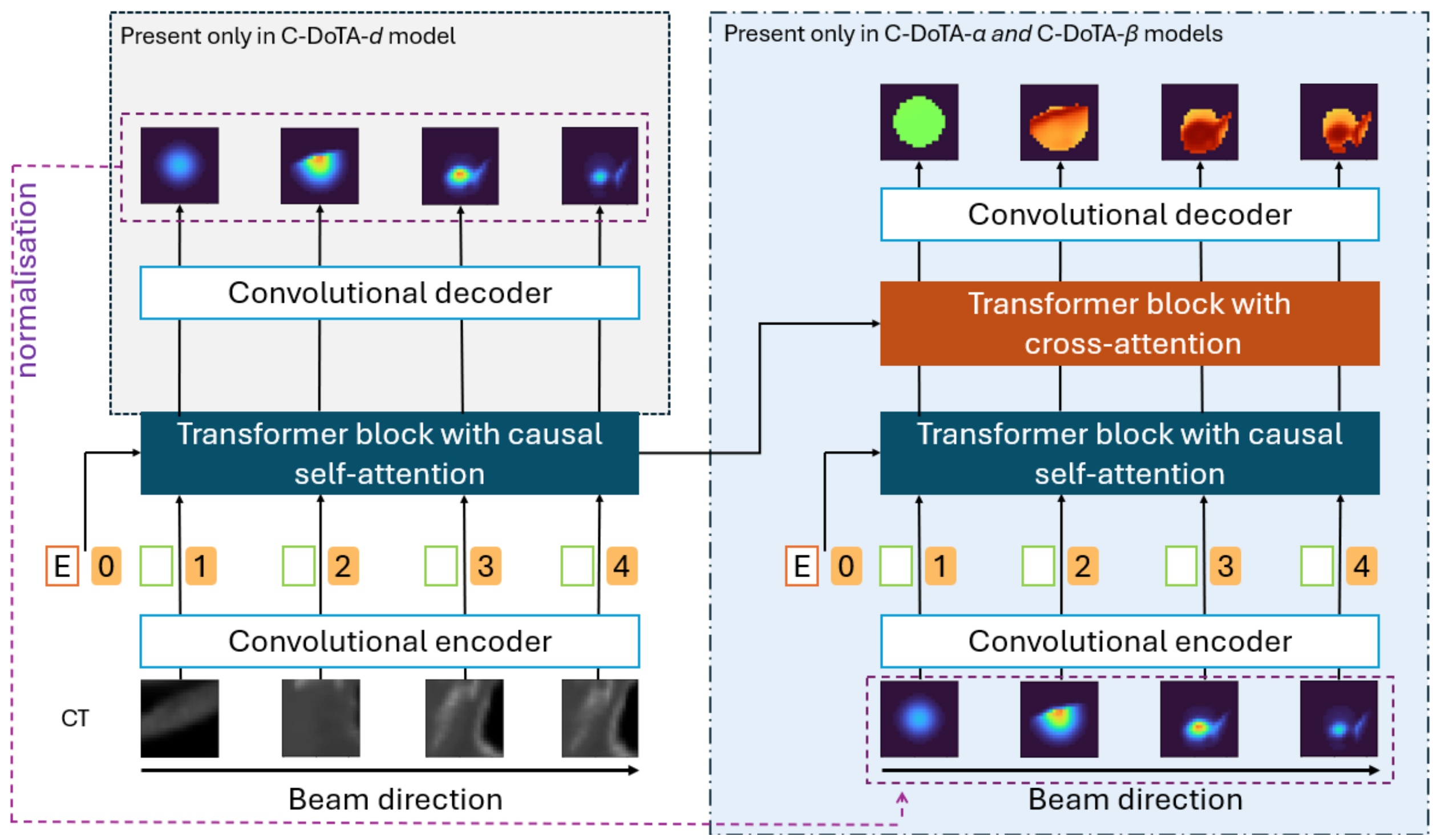

*Figure 1 C-DoTA architectures. Left: Vanilla DoTA, with a convolutional encoder passing the normalized BEV-CT as a sequence of 2D images to the transformer block, together with the normalised energy E, used for dose prediction (C-DoTA-d). Right: The architecture for predicting α and β parameters is expanded by an additional convolutional encoder and transformer with the predicted dose as an input. The output of both transformers for CT and predicted dose is finally processed by a third transformer block using cross attention.*

## 2.4.1 Data Preprocessing

Each PB simulated in FRED was cropped symmetrically to 24x24x240 $mm^3$. To prevent training on the noise at the beam penumbra, dose values below a given threshold were masked and zeroed out. For training of C-DoTA-d, this threshold was defined as 1% of the sample maximum in the FRED-computed absorbed dose. For consistency between generated training data and the intended application of the model, for C-DoTA-α and C-DoTA-β, a threshold of 1% of the sample maximum from the C-DoTA-d prediction was used instead (as during inference only the C-DoTA-d predicted is available). For distinction, the two masks are called “MC mask” and “AI mask” for the remainder of the document. This workflow is illustrated in Figure S3.

Unlike the vanilla DoTA model, the C-DoTA models normalised only the inputs, such as CT data and energy and the absorbed dose predicted by C-DoTA-d, range to [0,1]. Normalisation of the labels caused blurry output, especially near the Bragg peak. Instead, we used the following FRED labels in our loss function: absolute absorbed dose values, alpha scaled by 10 and beta values by $10^3$. This resulted in value ranges for the absorbed dose of [0 – 2.6 Gy], for alpha [0 -15.8 $Gy^{-1}$], and for beta [0 – 26.5 $Gy^{-2}$]. This scaling increased the values of the loss function and enabled more effective updates of the trainable weights. Moreover, interpretable thresholds could be employed in the custom loss, see the following section.

### 2.4.2 Model training details

The custom loss function was designed based on MSE with three additional components, where $y$ is the GT value and $\tilde{y}$ is the predicted value:

**Masked MSE**:

$$Masked\ MSE\ = \frac{1}{n} \sum_{i:\ |y_i - \tilde{y}_i| > \epsilon^2} (y_i - \tilde{y}_i)^2 \tag{0.3}$$

where n is the cardinality of the set of voxels with a squared error above the threshold $\epsilon^2$ (0.001 Gy for dose, 1/Gy for alpha and 1/$Gy^2$ for beta) at the given epoch. Unlike the traditional MSE loss, this masked loss excludes a high number of voxels with very small differences, forcing the model to pay more attention to the regions with high squared dose error, i.e., emphasising the narrow Bragg Peak where gradients are high.

**Depth-wise partial gradient loss:**

$$Depth\ Loss\ = \frac{1}{m} \sum_{i}^{m} \left( \frac{\partial y_i}{\partial z} - \frac{\partial \tilde{y}_i}{\partial z} \right)^2 \tag{0.4}$$

where the partial derivatives of the true and predicted responses with respect to depth $z$ for each voxel $i$ are calculated as

$$\frac{\partial y_i}{\partial z} = \frac{y_{j:\ z_j = z_i + \Delta z}\ -\ y_i}{\Delta z} \tag{0.5}$$

$$\frac{\partial \tilde{y}_i}{\partial z} = \frac{\tilde{y}_{\ j:\ z_j = z_i + \Delta z} - \tilde{y}_i}{\Delta z}, \tag{0.6}$$

with $j$ being the index of the voxel in the next depth layer adjacent to voxel $i$ , $\Delta z$ is the depth spacing and $m$ is the total number of voxels.

**Region masked MSE:**

$$Region\ masked\ MSE\ = \frac{1}{l} \sum_{i: y_i > \theta} (y_i - \tilde{y}_i)^2 \tag{0.7}$$

where $l$ is the cardinality of the set of voxels where GT values are above the threshold $\theta$ (60% of the sample maximum) at the given epoch. The loss increases the weight of the error in the regions with high values. These regions have the highest error if the model is trained on MSE only. The decision of the threshold was based on an empirical study.

The final loss is defined as the weighted sum of the separate MSE losses with the weight factors listed in Table S1:

$$Custom\ loss = MSE + w_{Mask} \cdot Masked\ MSE + w_{Depth} \cdot Depth\ Loss\ MSE + w_{Region} \cdot Region\ masked\ MSE \quad (0.8)$$

To find the optimal weight, we trained models with different sets of weights in the interval [0,1] and a step of 0.2. The optimal weight values were chosen based on the evaluation of the validation dataset. Not all the weights were used in all models.
The C-DoTA models were first trained for two times 28 epochs using only a mean squared error (MSE) loss, similar to the vanilla model. This was followed by two more times 28 epochs using a custom loss. The learning rate was rescheduled after each block of 28 epochs to prevent convergence in a local minimum.

## 2.5 *Evaluation*

As evaluation metrics, we used the GPR with a 1%/1 mm criteria and a 10% threshold. The 10% threshold was applied to mitigate the influence of the metric in low-dose regions, as low values artificially inflated the GPR. In the results, the GPR was computed only for voxels where the C-DoTA-d prediction exceeded 1% of its maximum, i.e., within the AI dose mask. This was necessary due to the impact of the AI dose mask on the alpha and beta predictions. The GPR based on the MC mask for the MC label can be found in the supplementary materials.

Furthermore, the consistency between prediction and GT was checked with dice coefficients [27]

by calculating iso-dose contours for C-DoTA-d at 1%, 30% and 70% of the sample maximum. The maximum of the GT sample was used to calculate relative metrics such as MSE. To estimate the impact of voxel exclusion due to differences in the 1% dose level between AI and MC, we counted voxels present in the MC mask but not in the AI dose mask and vice versa.

In addition to 3D-dose analyses, integrated depth-dose (IDD) profiles were assessed. Unlike GPR for IDD metrics, the MC mask was applied to GT, and the AI dose mask to the AI output to calculate the carbon range R80 of the Bragg peak, relative dose difference (ΔDmax) and MSE.

To automatically compute the carbon range R80 of the Bragg peak, the Python Shapely package was used to find the interpolated intersection value at 80% of the sample maximum. This allowed us to evaluate the range accuracy of the model dose output.

To estimate the model uncertainty distribution, we applied Monte Carlo dropout (MCD) [28]. MCD is an efficient method for estimating the stability of a model's performance through multiple forward inferences with an activated dropout layer, thereby simulating multiple trained models. The evaluation was performed with a dropout rate of 0.2 on the complete prediction chain (and the dropout layers for both training and inference were the same as in the vanilla DoTA model).

In our setup, the alpha and beta predictions are strongly correlated to the quality of the C-DoTA-d dose input. In this work, we evaluated the propagated uncertainty using the MCD C-DoTA-d outputs performed with activated dropout layers. It means that the output of C-DoTA-d is used as input for C-DoTA-α and C-DoTA-β models. After 30 runs, the average value and standard deviation per sample were analysed. For the graphic representation of the MCD-based uncertainty, a standard dose prediction was used as input to show the uncertainty regions of the models only. Quantitative statements on model uncertainty were obtained by evaluating the gamma passing rate of the 30 individual predictions.

# 3 Results

Table 1 shows the results of GPR 1%/1 mm with the 10% threshold applied to the output of all three C-DoTA models (absorbed dose (D), alpha (α) and beta (β)). The median value for all three outputs is around 99%. 5%-tile values are 98.09% for absorbed dose, 95.60% for alpha and 95.73% for beta, which shows that most of the data is close to the expected values. The results for the MC mask are presented in the Supplementary Materials (Table S2).

*Table 1 GPR 1%/1 mm results. AI dose mask was applied to both GT and AI output.*

| Parameter | Value range | GPR (1%/1 mm) [%] | | | | |
|---|---|---|---|---|---|---|
| | | Median | Min | 10%-tile | 90%-tile | Max |
| D | 0-2.6 Gy | 99.76 | 92.06 | 98.70 | 99.96 | 100 |
| α | 0-1.6 $Gy^{-1}$ | 99.14 | 86.36 | 96.72 | 99.87 | 100 |
| β | 0-0.03 Gy-$^{2}$ | 98.74 | 85.26 | 96.71 | 99.45 | 99.87 |

The top of Figure 2 shows the output of the best absorbed dose prediction with GPR 1%/1 mm of 100% as well as the corresponding alpha (100%) and beta (99.87%) predictions. In contrast, the bottom one shows the difference between GT and the prediction, as well as the corresponding MCD uncertainty maps. The results with MC mask applied to GT is showed in Figure S4. The top of Figure 3 shows a test sample with the worst GPR for physical dose prediction, at 92.06%, with alpha at 89.16% and beta at 90.62%. The bottom part illustrates its difference and MCD uncertainty. The worst-case sample has a highly inhomogeneous structure, featuring bones and a complex nasal cavity, where the Bragg peak is positioned exactly on the boundary between the regions of high and low density, resulting in a complex shape of the beamlet. The main uncertainty region for the dose is around the Bragg peak region for all the models, where the gradient is sharp. This somewhat correlates with the difference in the best-case scenario; however, no such relation is seen in the worst-case scenario. The results with MC mask applied to GT is showed in Figure S5.

The worst case of alpha (Figure S6) and beta (Figure S7) predictions can be found in the supplementary materials.

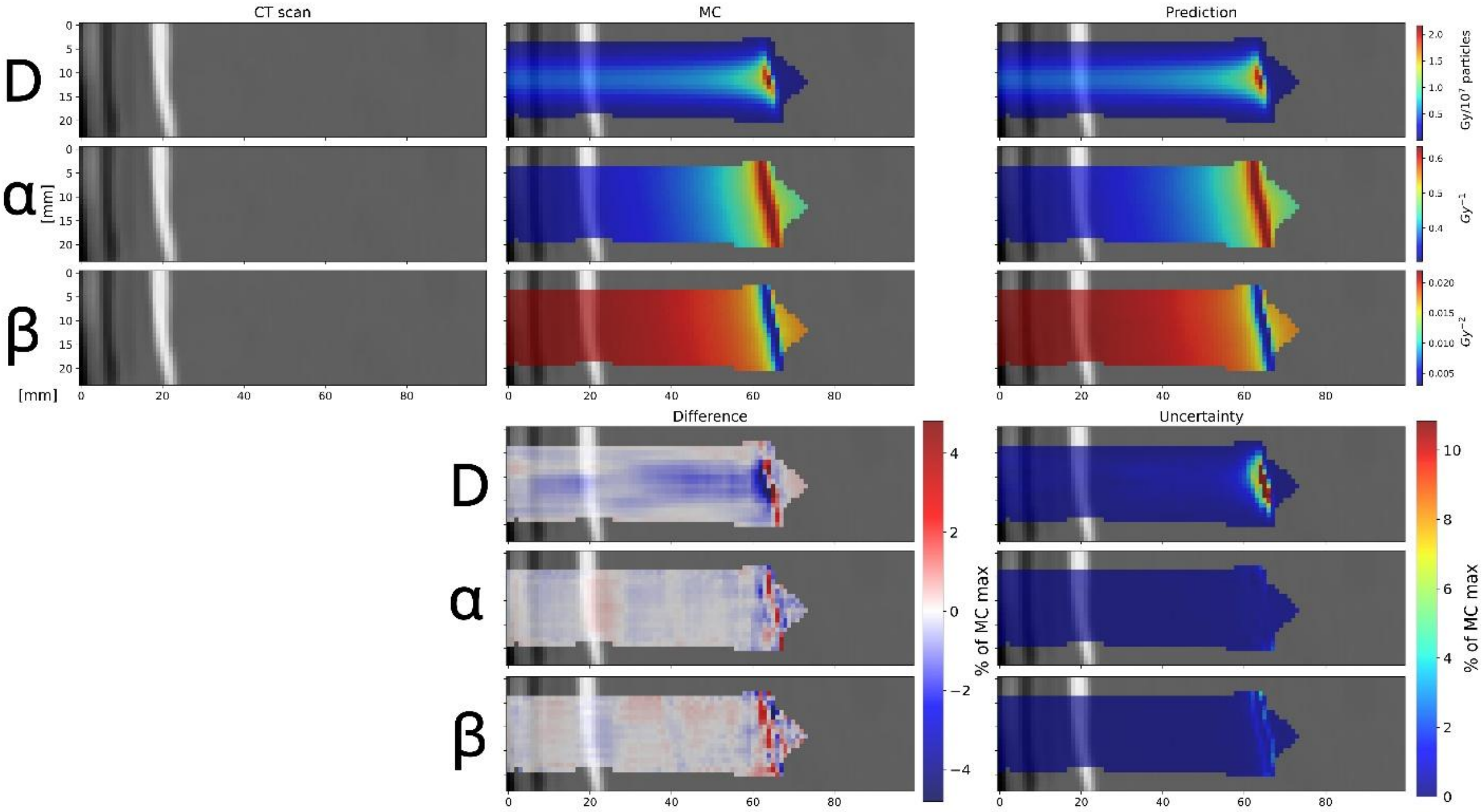

*Figure 2 Best result in terms of the GPR for the physical dose. The GPR with a 1%/1 mm was top row absorbed dose (D): 100%, middle row alpha (α): 100%, bottom row beta (β): 99.87%. The top part: on the left, a slice of the BEV CT is shown; in the centre, the MC GT is shown; and the C-DoTA predictions on the right. The bottom part: on the left side, the difference (MC minus C-Dota); on the right side, MCD uncertainty. For better visibility, the maximum and minimum values of the colour scale in the difference plots were set to the maximum between the absolute values of the 99.9%-tile and the 0.1%-tile.*

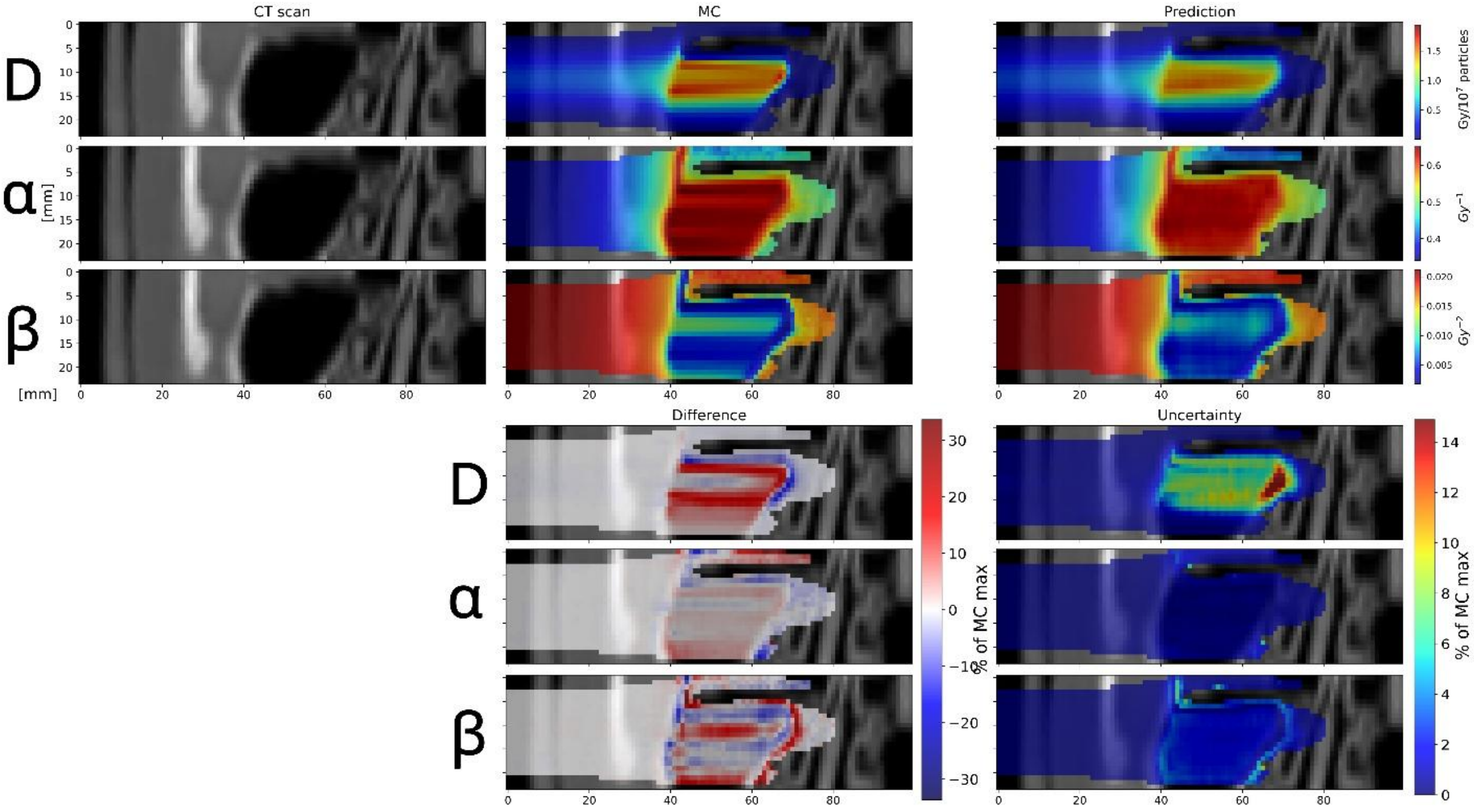

*Figure 3 Worst result in terms of the GPR for the physical dose. The GPR with a 1%/1 mm was top row absorbed dose (D): 92.06%, middle row alpha (α): 89.16%, bottom row beta (β): 90.62%. The top part: on the left, a slice of the BEV CT is shown; in the centre, the MC GT is shown; and the C-DoTA predictions on the right. The bottom part: on the left side, the difference (MC minus C-DoTA); on the right side, MCD uncertainty. For better visibility, the maximum and minimum*

*values of the colour scale in the difference plots were set to the maximum between the absolute values of the 99.9%-tile and the 0.1%-tile.*

To illustrate the effect of the custom loss, the results of the GPR 1%/1 mm for the models trained for 56 and 112 epochs on MSE can be found in the Supplementary materials (

Table S3 and Table S4). While the physical dose prediction does not benefit much in GPR, the alpha and beta have gained ≈18% in minimum values after 112 epochs. The additional 56 epochs improved the GPR by further ≈3% at minimum.

The result of the MCD uncertainty prediction is summarised in Table 2 in the form of the metrics for the GPRs evaluated for each of the individual 30 MCD predictions. The median of the mean standard deviation was below 0.5% for all the models, which shows high reliability of the prediction.

*Table 2 Mean (top row) and standard deviation (bottom row) of GPR per sample after 30 runs with MCD of 0.2. AI dose mask was applied to both GT and AI output.*

| **Parameter** | **Value range** | **GPR (1%/1 mm) [%]** | | | | |
|---|---|---|---|---|---|---|
| | | Median | Min | 10%-tile | 90%-tile | Max |
| D | 0-2.6 Gy | 99.59 | 92.62 | 96.62 | 99.89 | 100 |
| | | 0.18 | 0.01 | 0.06 | 0.67 | 3.49 |
| α | 0-1.6 $Gy^{-1}$ | 98.62 | 86.97 | 96.42 | 99.57 | 99.96 |
| | | 0.47 | 0.04 | 0.15 | 1.52 | 6.28 |
| β | 0-0.03 $Gy^{-2}$ | 98.20 | 85.59 | 96.12 | 99.05 | 99.52 |
| | | 0.39 | 0.05 | 0.16 | 1.42 | 9.57 |

Table 3 presents a more detailed analysis of the test dataset. The mode of the range difference is almost 0 mm, while the automatic detection of R80 causes some extreme differences. The median relative difference in the dose maximum is 0.9%. The RMSE indicates that the average difference between the GT and prediction is 0.63%, or cumulatively, 0.62 Gy/$10^7$ primaries per 1 mm depth.

Examples of the extremes are shown in Figure S8. The left plot depicts the maximum difference in R80, while the middle one shows the minimum difference. It is clearly visible that the predicted IDD profile closely resembles the MC one, but due to its complex shape, the R80 value obtained is highly sensitive to minor changes, in this case causing a shift between the first and second peak in the IDD profile. In the right plot, the dose loss in the Bragg peak region is caused by a dense, bony CT structure, which also poses a challenge for PBAs to calculate. The corresponding C-DoTA outputs can be found in Figure S9.

The evaluation of the Dice coefficient between MC and C-DoTA predictions indicates a high level of agreement between the isodose lines. A small number of voxels primarily caused low values in some samples, where the relative impact of a small error is greater. The 10 and 90 percentile values are close to the median, demonstrating the stability of the C-DoTA results.

There is also a fraction of voxels that, due to the threshold, are present in GT but not in the prediction, and vice versa (Table S5). This fraction of the lost voxels due to thresholding is 8% with mean of 0.02 Gy/$10^7$ primaries and a standard deviation of 0.003 Gy/$10^7$ primaries.

As an additional test, not included in the tables, we checked the behaviour of our models on some extreme cases such as a lot of air in the nasal cavities due to a possible surgery before the therapy or for beam paths that pass many bone-tissue interfaces. This data is presented in the supplementary material to this manuscript (Figure S10).

Finally, we analysed the speed of the C-DoTA dose engine versus the FRED GPU MC. The time measurements were performed on a GPU Node with 8 AMD Radeon Instinct Mi100 32 GB video cards, 96 CPUs and 384 GB RAM. Only 1 out of 8 GPUs was used, and the whole node was still blocked for other users to obtain a clean measurement of the runtime. The time for C-DoTA, including loading the samples into the model and dose pre-processing for alpha and beta models (but not the interpolation for obtaining the BEV CT), was 0.032±0.001 s (0.007±0.0002 s for the absorbed dose only). The samples were randomly selected from the test dataset. C-DoTA is capable of predicting >400 times faster than MC FRED (14 s) for our setup.

*Table 3 Additional evaluation metrics. *values calculated using the AI dose mask*

| | | **Median** | **Min** | **10%-tile** | **90%-tile** | **Max** |
|---|---|---|---|---|---|---|
| IDD Curve | ΔR80 mm | -0.02 | -24.48 | -0.44 | 0.32 | 26.20 |
| | ΔDmax % | 0.90 | -11.51 | -2.25 | 4.32 | 18.38 |
| | RMSE %* | 0.97 | 0.25 | 0.65 | 1.61 | 6.12 |
| FRED Vs. C-Dota | RMSE %* | 0.14 | 0.05 | 0.09 | 0.23 | 0.63 |
| | Dice (1% isodose)<br>(n voxels in GT) | 0.95<br>(19241) | 0.84<br>(11531) | 0.93<br>(14455.5) | 0.96<br>(24566) | 0.98<br>(38586) |
| | Dice (30% isodose)<br>(n voxels in GT) | 0.90<br>(640) | 0.29<br>(301) | 0.77<br>(418) | 0.96<br>(2016.5) | 0.99<br>(5907) |
| | Dice (70% isodose)<br>(n voxels in GT) | 0.83<br>(39) | 0.16<br>(15) | 0.68<br>(29) | 0.92<br>(77) | 1<br>(499) |

# 4 Discussion

This work presents the first AI-based dose engine for carbon ion therapy capable of predicting both absorbed dose and the α and β parameters required for RBE-weighted dose calculation and optimisation. The high GPRs (>98% median) and strong spatial agreement (Dice coefficient 0.95 at 1% isodose) demonstrate that the proposed framework delivers robust predictions across a wide range of anatomical configurations. Accuracy loss in highly heterogeneous regions—also a challenge for PBAs—was modest and primarily affected the high-gradient dose falloff. These cases highlight the importance of targeted data augmentation or hybrid MC–AI workflows to further improve performance in extreme anatomies, such as post-surgical cavities or dense bony structures.

Clinically, the model's independence from simulation statistics, beam energy and anatomy for inference time makes it well-suited for online adaptive planning and 4D robust optimisation. Parallelisation across GPUs could potentially reduce runtime to milliseconds for full plans, enabling real-time replanning during treatment sessions. The high quality PB data output of the model can be directly used for optimisation, as opposed to other MC [29] or AI models [30] that compute or predict the full dose of a plan.

However, several limitations must be acknowledged. The training data were restricted to head-and-neck patients, a fixed horizontal beamline, and a therapeutic energy range of 115–260 MeV/u. A broader generalisation will require expanding the dataset to include more anatomical sites, beam geometries, and beamlines. Additionally, while α and β predictions matched MC values, their biological accuracy depends on the local effect model and requires further validation against experimental radiobiology data. Future research should address the following issues: the effect of the absence of the low-dose region in the C-DoTA prediction (8% of voxels on average) and the error introduced by alpha and beta predictions on RBE-weighted dose, both of which need to be investigated in the context of a full plan. This full plan can also be compared against another MC model, featuring a more complete physics implementation at longer calculation times, to benchmark the simplifications inherent to FRED's GPU implementation. Moreover, the consistency between the base data for optimisation and dose calculation reduces the error of the final plans. Finally, to generalise the model for various treatment sites, the inclusion of tissue-specific alpha/beta ratios needs to be studied. FRED is a GPU-based MC tool for fast simulation and includes accurate representation of the beam transport and interaction at voxel level delivering the most important characteristics needed to train the C-DoTA model. Nevertheless, it is not a clinically tested tool and does not consider refragmentation. For clinical application the models need to be fine-tuned on the data created with more accurate and currently slower MC tools.

Overall, this proof-of-concept demonstrates that deep learning can deliver input parameters for biologically-weighted dose needed for carbon ion therapy with clinical runtimes and MC-quality,, paving the way for integration into routine workflows and wider adoption of this modality.

# 5 Conclusions

We developed an AI model able to calculate the input parameters for RBE-weighted dose calculation in carbon-ion therapy with MC accuracy but at 400-fold faster computation time. The model's speed can potentially enable rapid online re-planning, which is particularly valuable for addressing inter-fractional anatomical changes. By removing the need for MC, C-DoTA will improve carbon ion therapy accuracy, enabling safer and more effective treatments. Future directions include lowering the prediction threshold to 0.5% (clinically required); collaboration with C-ion clinical centres to test robustness across beamlines and patient cohorts; and embedding the model into clinical workflows (e.g., TRiP98) to evaluate end-to-end plan quality.

# Conflict of interest

Zoltan Perko is an associate professor at Delft University of Technology and works as Senior Applied Scientist at Radformation Inc. His industry employment has no connection to the presented work. The other authors declare that they have no known competing financial interests or personal relationships that could have appeared to influence the work reported in this paper.

# Supplementary materials

*Table S1 Weights used in the custom loss function*

| Parameter | $w_{Mask}$ | $w_{Depth}$ | $w_{Region}$ |
|---|---|---|---|
| C-DoTA-d | 0.4 | 1 | 0 |
| C-DoTA- α | 0 | 0 | 1 |
| C-DoTA- β | 0 | 0 | 0.4 |

*Table S2 GPR 1%/1 mm results computed on the MC mask*

| Parameter | Value range | GPR (1%/1 mm) [%] | | | | |
|---|---|---|---|---|---|---|
| | | Median | Min | 10%-tile | 90%-tile | Max |
| D | 0-2.5 Gy | 99.76 | 92.06 | 98.70 | 99.96 | 100 |
| α | 0-1.6 $Gy^{-1}$ | 91.17 | 71.49 | 87.28 | 93.77 | 97.58 |
| β | 0-0.03 $Gy^{-2}$ | 91.37 | 71.69 | 87.66 | 93.80 | 97.69 |

*Table S3 GPR 1%/1 mm results after 56 epochs with MSE. AI dose mask was applied to both GT and AI output.*

| Parameter | Value range | GPR (1%/1 mm) [%] | | | | |
|---|---|---|---|---|---|---|
| | | Median | Min | 10%-tile | 90%-tile | Max |
| D | 0-2.5 Gy | 99.61 | 88.66 | 98.02 | 99.94 | 100 |
| α | 0-1.6 $Gy^{-1}$ | 95.76 | 64.53 | 88.80 | 98.44 | 99.91 |
| β | 0-0.03 $Gy^{-2}$ | 95.39 | 62.65 | 87.90 | 98.20 | 99.61 |

*Table S4 GPR 1%/1 mm results after 112 epochs with MSE. AI dose mask was applied to both GT and AI output.*

| Parameter | Value range | GPR (1%/1 mm) [%] | | | | |
|---|---|---|---|---|---|---|
| | | Median | Min | 10%-tile | 90%-tile | Max |
| D | 0-2.5 Gy | 99.66 | 91.54 | 98.46 | 99.94 | 100 |
| α | 0-1.6 Gy$^{-1}$ | 97.28 | 69.61 | 92.25 | 98.99 | 99.89 |
| β | 0-0.03 Gy-$^{2}$ | 96.26 | 65.93 | 90.61 | 98.56 | 99.73 |

*Table S5 Additional evaluation metrics for voxel-wise comparison between FRED MC and C-DoTA outputs to show the impact of the two masks.*

| | | Median | Min | 10%-tile | 90%-tile | Max |
|---|---|---|---|---|---|---|
| Voxels present in GT and not present in the prediction | Count | 1479 | 292 | 983 | 2416 | 8530 |
| Voxels present in GT and not present in the prediction | (in %) | 8 | 1.32 | 5.48 | 11.42 | 23.38 |
| Voxels present in GT and not present in the prediction | Total voxels in GT | 19240 | 11531 | 14455 | 24566 | 38586 |
| Voxels present in GT and not present in the prediction | Mean, Gy/$10^7$ primaries | 0.02 | 0.01 | 0.02 | 0.02 | 0.03 |
| Voxels present in GT and not present in the prediction | Mean± Standard deviation, % | 1.17±0.17 | 1.08±0.07 | 1.14±0.12 | 1.25±0.29 | 1.83±1.31 |
| Voxels present in the prediction and not present in GT | count | 206 | 1 | 71 | 690 | 6973 |
| Voxels present in the prediction and not present in GT | (in %) | 1.15 | 0.00 | 0.37 | 4.23 | 21.09 |
| Voxels present in the prediction and not present in GT | Total voxels in prediction | 18032 | 10669 | 13577 | 22817 | 36523 |
| Voxels present in the prediction and not present in GT | Mean, Gy/$10^7$ primaries | 0.02 | 0.01 | 0.02 | 0.03 | 0.05 |
| Voxels present in the prediction and not present in GT | Mean± Standard deviation, % | 1.33±0.40 | 0.97±0.00 | 1.15±0.18 | 1.61±0.86 | 2.46±2.57 |

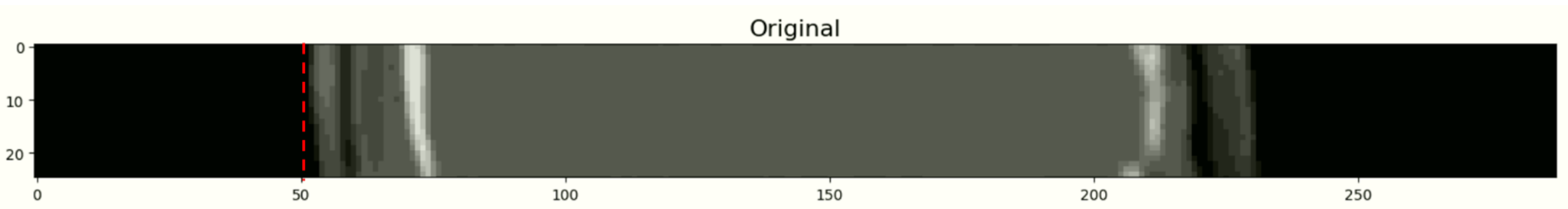

*Figure S1 Example of a BEV-CT cut along the beam axis impinging from the left. The red dashed line indicates the cut of the proximal slices containing only air.*

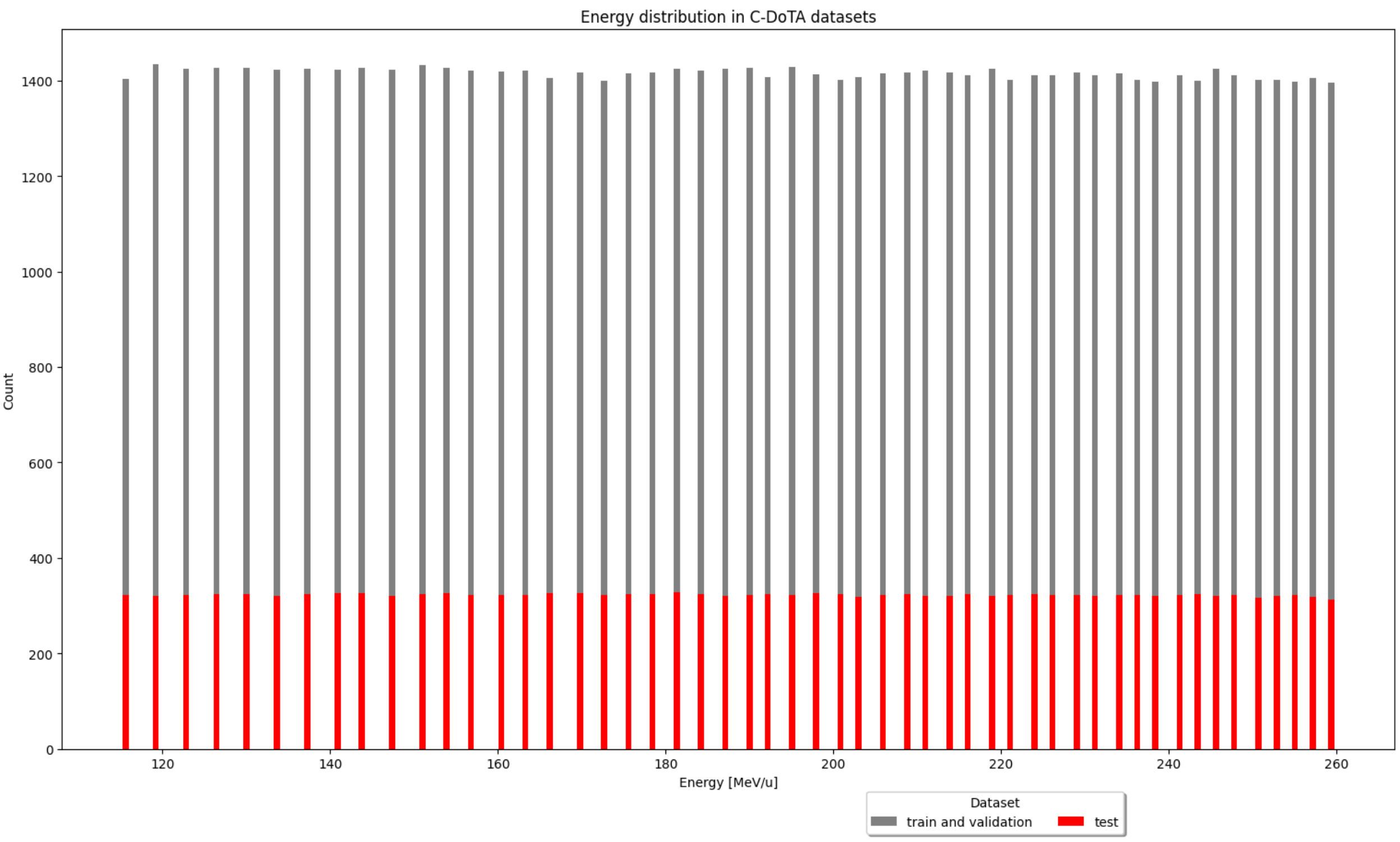

*Figure S2 The number of samples in C-DoTA database for discrete energies used at CNAO facility, Italy.*

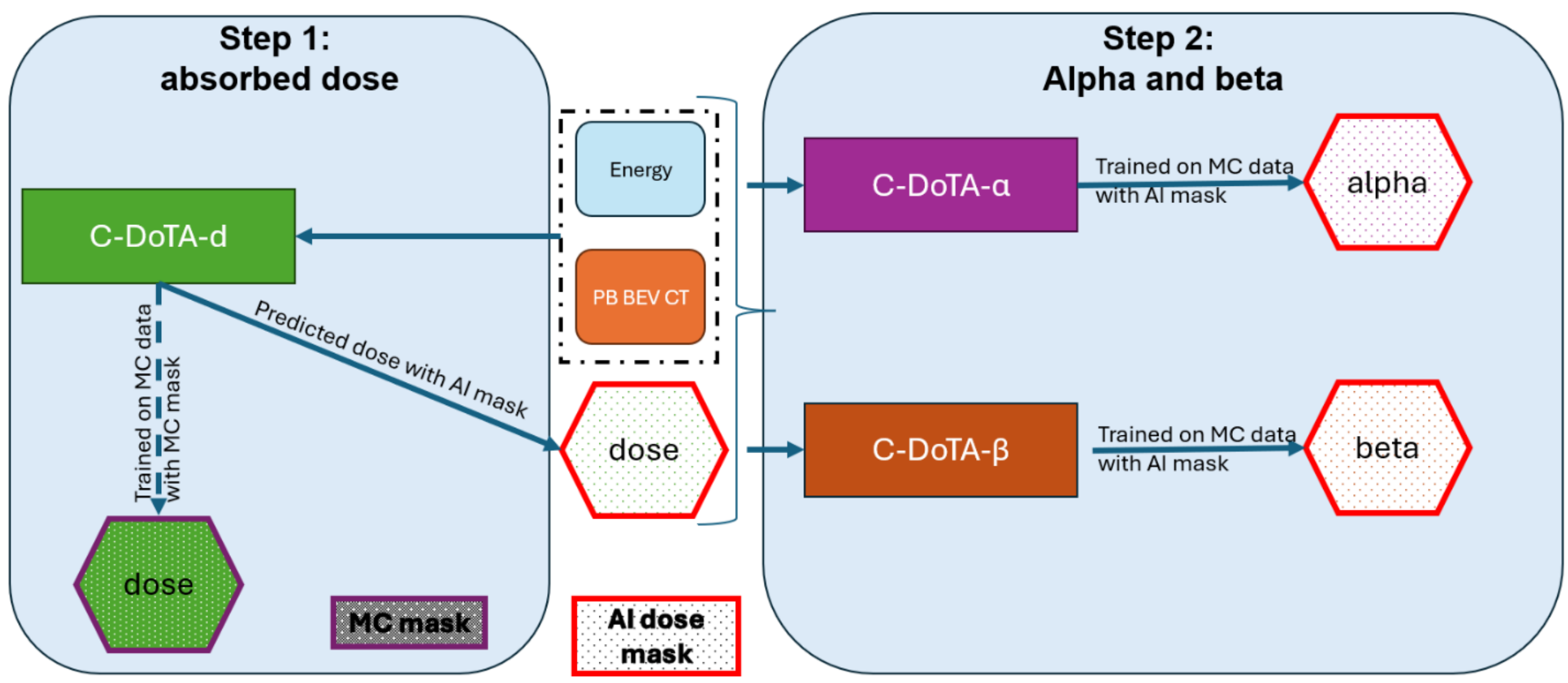

*Figure S3 Data workflow. The output of the C-DoTA-d model is normalised to be used as an input.*

The following Figures S4-7 and S9, S10 have the same structure: on the left, a longitudinal slice of the BEV CT is shown, followed by MC-GT and the C-DoTA prediction. The final column shows the difference of MC minus C-DoTA. For better visibility, the maximum and minimum values of the colour scale of the difference plots were set to the maximum between the absolute values of the 99.9%-tile and 0.1%-tile.

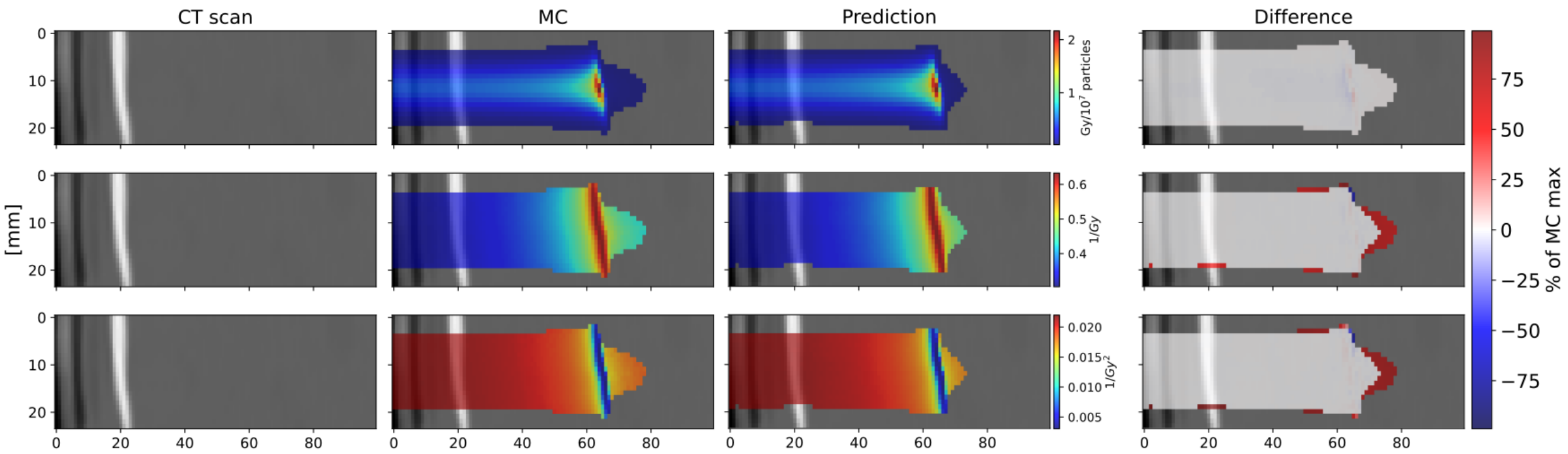

*Figure S4 Best physical dose GPR result. To illustrate the difference between the applied masks, the MC results employ the MC mask, the C-DoTA results employ the AI mask. The difference shows all values without considering the masks. GPR 1%/1 mm: top row, dose:100%; middle row, alpha:100%; bottom row, beta: 99.87%*

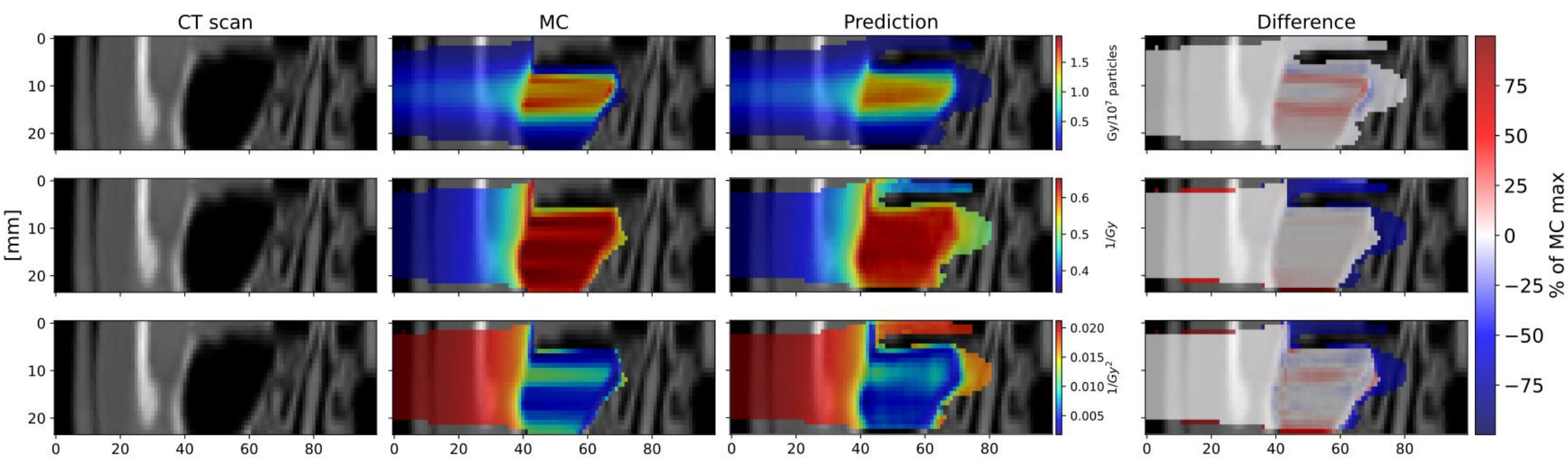

*Figure S5 Worst physical dose GPR result. To illustrate the difference between the applied masks, the MC results employ the MC mask, the C-DoTA results employ the AI mask. The*

*difference shows all values without considering the masks. GPR 1%/1 mm: top row, dose :92.06%; middle row, alpha: 89.16%; bottom row, beta: 90.62%*

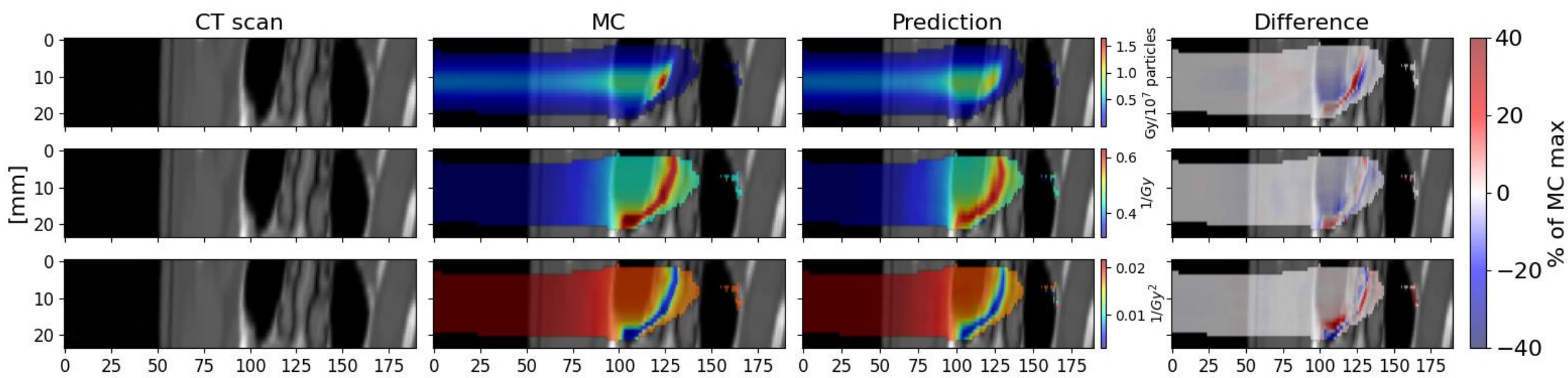

*Figure S6 Worst result GPR for alpha. GPR 1%/1 mm: top row, dose: 95.28%; middle row, alpha: 86.36%; bottom row, beta: 87.76%.*

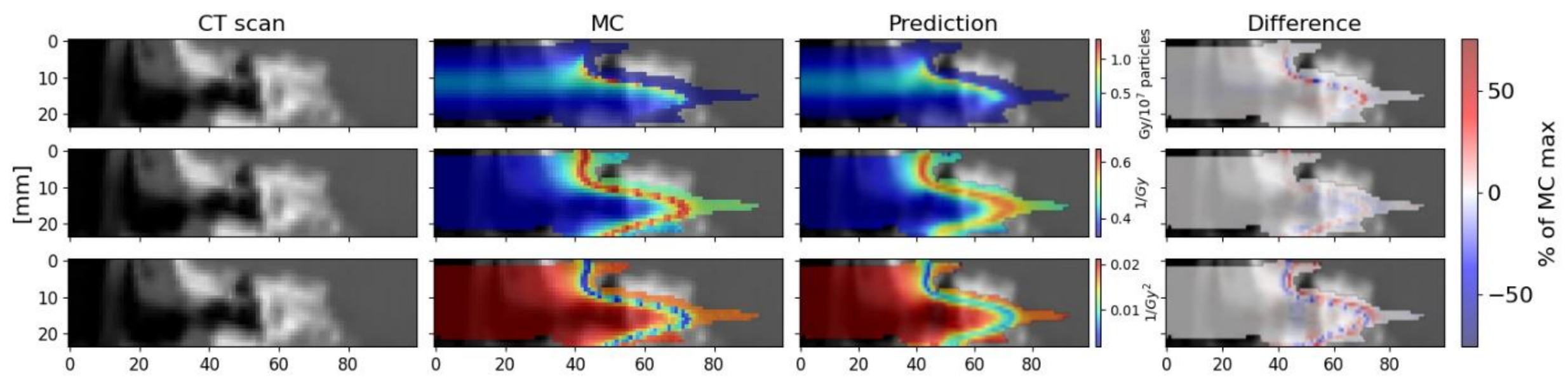

*Figure S7 Worst result GPR for beta. GPR 1%/1 mm: top row, dose: 94.53%; middle row, alpha: 88.73%; bottom row, beta: 85.26%*

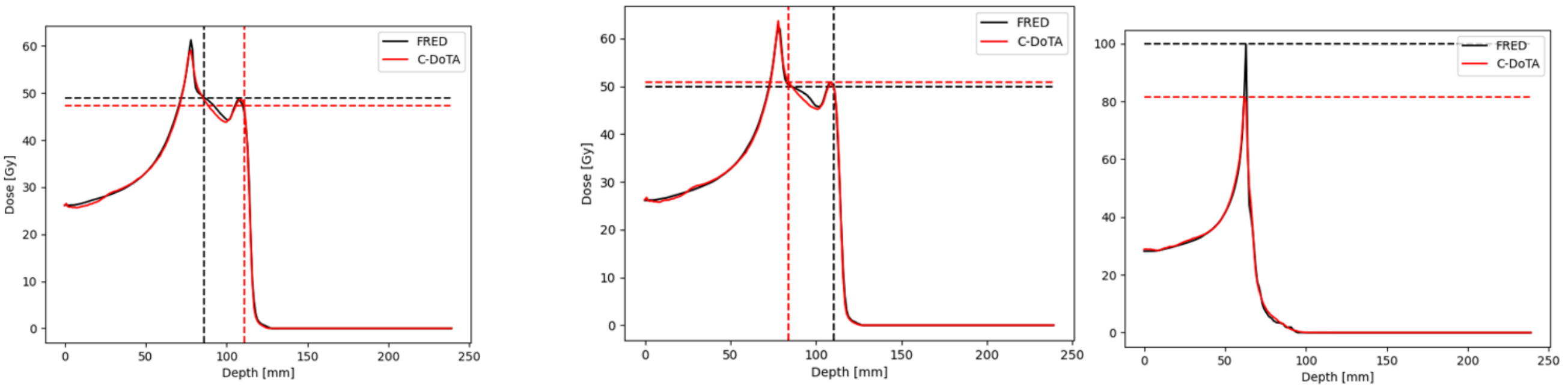

*Figure S8 IDD Curve comparison. The left figure corresponds to ΔR80 min, the middle one ΔR80 max and the right one to the maximum of ΔDmax*

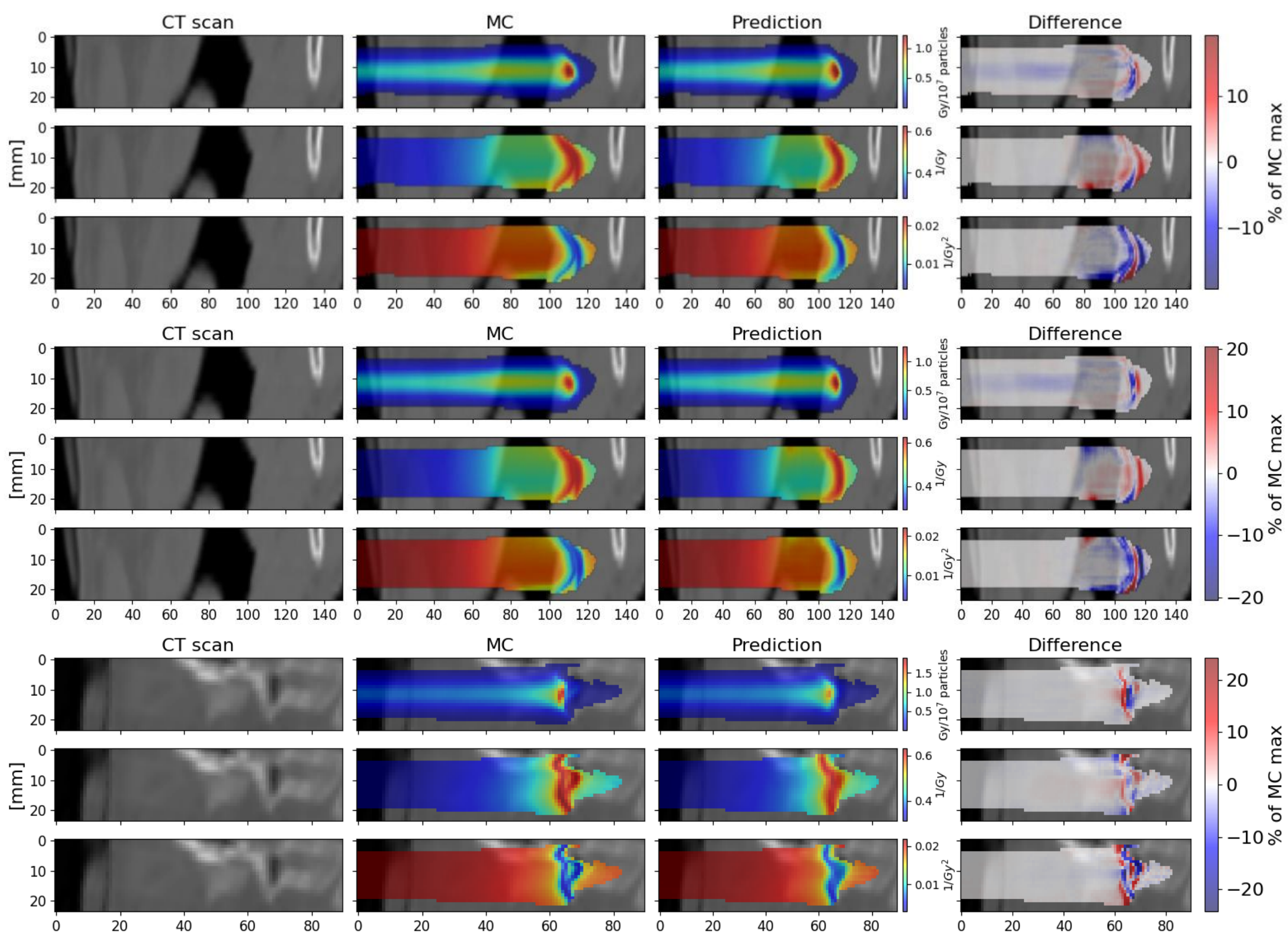

*Figure S9 The top row corresponds to Figure S8 left panel, the middle row to the middle panel and the bottom row to the right panel.*

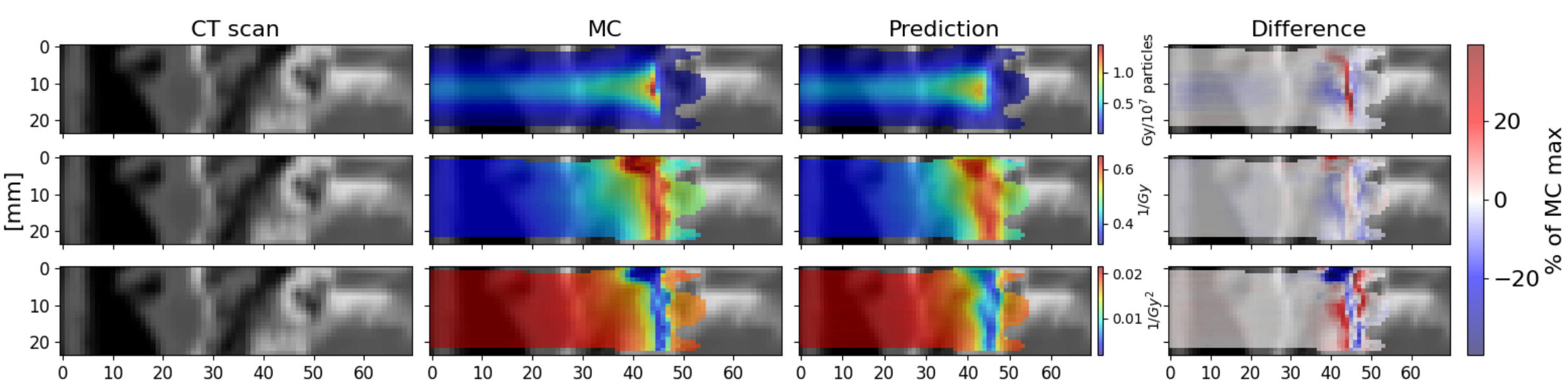

*Figure S10 An additional test of the models on an inhomogeneous structure. GPR 1%/1 mm: top row, dose: 94.60%; middle row, alpha: 87.42%; bottom row, beta: 80.26%*